\documentclass[aps,pra,twocolumn,superscriptaddress,showkeys,amsmath,amssymb,longbibliography]{revtex4-1}		
\usepackage{graphicx}
\usepackage{dcolumn}
\usepackage{bm}
\usepackage[colorlinks,urlcolor=blue,citecolor=blue,linkcolor=blue]{hyperref}
\usepackage{subfig}
\usepackage[font=small,labelfont=bf,format=plain,justification=centerlast,labelsep=period]{caption}
\usepackage[usenames, dvipsnames]{color}

\usepackage{verbatim}
\usepackage[normalem]{ulem}

\newcommand{\revision}[1]{\textcolor{black}{#1}}

\begin{document}

\preprint{APS/123-QED}

\title{Spectral signatures of non-thermal baths in quantum thermalization}

\author{Ricardo Rom\'an-Ancheyta}
\email{ancheyta6@gmail.com}
\address{Instituto Nacional de Astrof\'isica, \'Optica y Electr\'onica, Calle Luis Enrique Erro 1, Sta. Ma. \\ Tonantzintla, Puebla CP 72840, M\'exico}
\author{Bar{\i}\c{s} \c{C}akmak} 
\affiliation{College of Engineering and Natural Sciences, Bah\c{c}e\c{s}ehir University, Be\c{s}ikta\c{s}, Istanbul 34353, Turkey}
\author{\"Ozg\"ur E. M\"ustecapl{\i}o\u{g}lu}
\affiliation{Department of Physics, Ko\c{c} University, \.{I}stanbul, 
Sar{\i}yer, 34450, Turkey}
\date{\today}

\begin{abstract}
We show that certain coherences, termed as heat-exchange coherences, which contribute to the thermalization process of a quantum probe in a repeated interactions scheme, can modify the spectral response of the probe system. We suggest to use the power spectrum as a way to experimentally assess the apparent temperature of non-thermal atomic clusters carrying such coherences and also prove that it is useful to measure the corresponding thermalization time of the probe, assuming some information is provided on the nature of the bath. We explore this idea in two examples in which the probe is assumed to be a single-qubit and a single-cavity field mode. Moreover, for the single-qubit case, we show how it is possible to perform a quantum simulation of resonance fluorescence using such repeated interactions scheme with clusters carrying different class of coherences.
\end{abstract}



\maketitle

\section{Introduction}

In the rapidly growing field of quantum thermodynamics~\cite{Janet2016,JPA_Goold,arXiv_Alicki,Deffner2019book} several new effects that could bend the rules~\cite{Merali2017} of the classical theory 
are being predicted. Especially, developments related with the performance of non-thermal quantum machines constitute a good example to these effects, with which one can extend the limits dictated by the classical thermodynamics by exploiting the quantumness of the bath that the system of interest is interacting~\cite{Scully2002,Scully2003,gerzontemp,Angsar,hardal,DenizLevels,QCtranPRE,singlemachine,CerenEntropy,BarisThermal,Lutz2014,Niedenzu2016,Niedenzu2018}. The distinguishing property of such non-thermal baths is that they possess some amount of quantum coherence, and as a result they can not be described by a thermal state. A natural question at this point would be to ask about the effects of the coherences in the bath on the dynamics of the central system while they are in contact. It has been shown that it is possible to fully characterize the types of coherences in a non-thermal bath. While some coherences have a displacing or squeezing effects, there are certain types of coherences that merely contribute to the thermalization of the system, which are dubbed as {\em heat-exchange coherences} (HECs)~\cite{CerenEntropy,BarisThermal,gerzontemp,Angsar}. Therefore, non-thermal baths that only contain HECs act as an effective heat bath for the probe at a certain {\em apparent} temperature. The apparent temperature was recently introduced in \cite{Latune_2019} and it is based on the expression of the heat flow between, in general, out-of-equilibrium quantum systems. If one of the systems can be described as a bath, the apparent temperature coincides with notion of local temperature for stationary non-equilibrium baths~\cite{Alicki_2015,Alicki_2014}.

How to assign an apparent temperature to these non-thermal baths and how to test their corresponding quantum thermodynamic predictions from the spectroscopic point of view are the questions that lie at the heart of this contribution. We present a method that relies on the power spectrum of the probe system to assess the apparent temperature of the bath for which it is possible to show that, with some previous information of the bath, one can separately identify the effects of a thermal and a non-thermal, coherent bath. Moreover, the information about the thermalization time of the probe can also be extracted from the power spectrum. The presented method is actually along the same lines with the recent efforts in quantum thermometry, which tries to estimate the temperature of an environment by quantum probes~\cite{dePasquale,Stace_PRA_2010,dePasquale1,dePasquale2,dePasquale3,arXiv_Sanpera,arXiv_Paris,NJP_Steve,QST_Steve,PRL_Correa,PRL_Hofer}. Specifically, the present approach resembles the one adopted in~\cite{arXiv_Paris}, in which the authors take advantage of the dephasing dynamics of a qubit that is embedded in an environment to assign a temperature to that environment. However, in this work, the environment with which the probe is interacting need not to be a thermal one and can contain certain coherences that appears as an apparent temperature to the probe.

\begin{figure}[t]
\includegraphics[width=8.5cm, height=3.5cm]{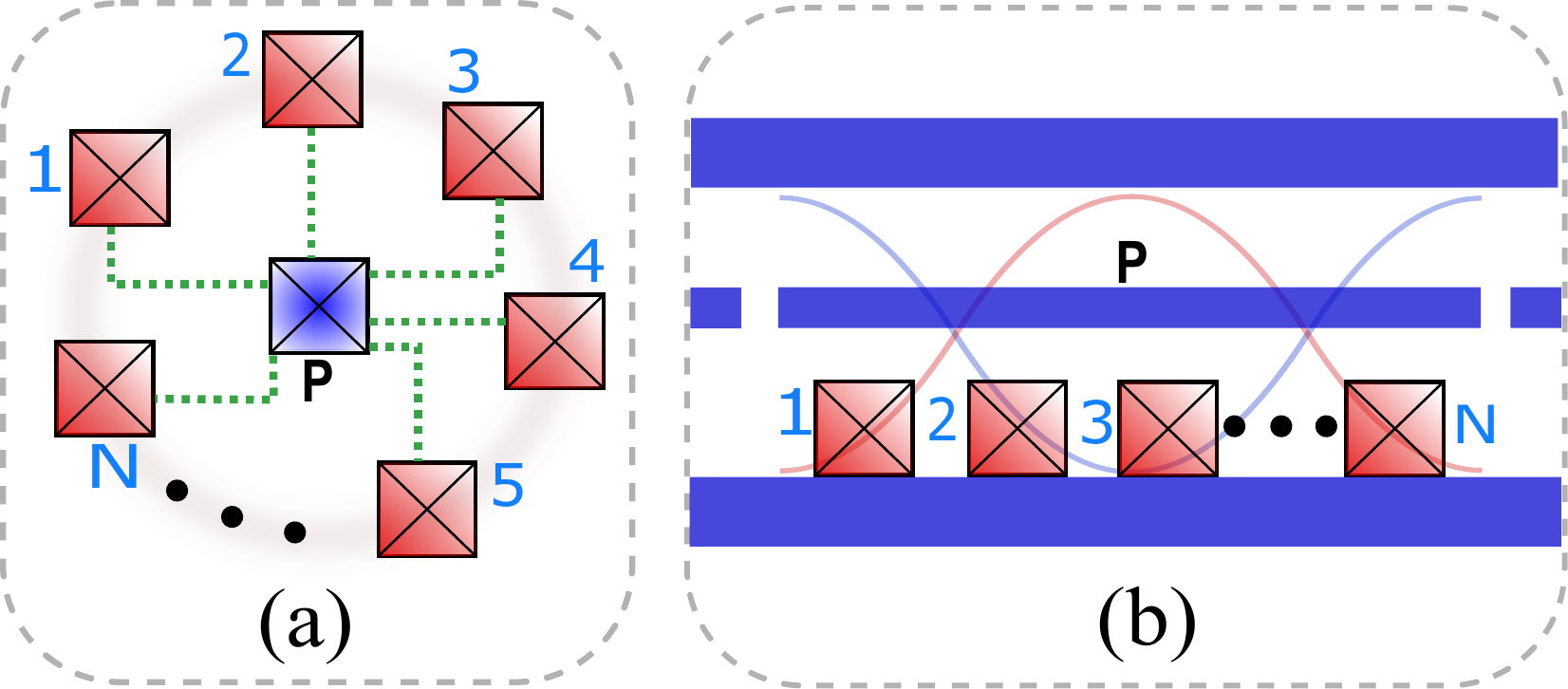} 
\caption{(Color online) Schematic representations 
of a repeated interaction schemes also known as generalized micromaser setups. 
(a)~A central probe qubit (blue central box) 
\textcolor{black}{interacts simultaneously}
with $N$-qubits 
(red surrounding boxes) 
acting as a non-thermal bath.
(b)~A single-electromagnetic field mode (blue thin 
rectangle) interacting collectively with a similar 
$N$-qubit cluster. These setups resemble the ones used in the field 
of circuit-QED, where boxes represent 
superconducting qubits and long rectangles are 
1D transmission line resonators. However, setups 
based on cavity-QED architecture where an atomic 
cluster beam traverse a 3D electromagnetic cavity will also work.\hspace{4.3cm}~}
\label{setup}
\end{figure}

In particular, we generalize the previous settings presented in~\cite{CerenEntropy,BarisThermal,gerzontemp,Angsar} and consider a probe system that is both in contact with a thermal and a non-thermal bath that only contain HECs, and address the aforementioned matter of thermalization dynamics by considering two simple, but not trivial, examples in the weak coupling regime. We begin by considering a central single-qubit as our probe [see Fig.~\ref{setup}(a)] which is interacting with a $N$-qubit coherent cluster (non-thermal bath) together with a thermal bath and derive some of their quantum thermodynamic effects on the probe, including the thermalization temperature and time. Then, we replace the qubit by a single electromagnetic field mode as our probe [see Fig.~\ref{setup}(b)], and we show that the apparent field temperature is the same as the qubit case, however, their respective thermalization times are completely different. Interestingly, when coherences that are different from HECs are considered to be present in the $N$-qubit cluster, this scheme can also be used to perform a quantum simulation of resonance fluorescence in an ordinary or squeezed vacuum in free space.

The rest of the manuscript is organized as follows. In Sec.~\ref{sec:model}, we introduce our model to manipulate the temperature of a quantum probe by interacting it with an $N$-qubit cluster that possess certain coherences. Sec.~\ref{sec:spectrum} presents our results on how to assess the apparent temperature of the coherent qubit clusters, and thermalization time of the quantum probe by looking at the stationary power spectrum of it. Also, we provide fundamental bounds on how well we can estimate the apparent temperature and the HECs of the non-thermal bath. In Sec.~\ref{dis_cohe}, by adopting a time-dependent physical spectrum, we identify which types of coherences can be used to shape the spectral response of the qubit probe and make a quantum simulation of resonance fluorescence. Sec.~\ref{sec:experiment} includes our discussion on the possible experimental applications of the proposed scheme and we conclude in Sec.~\ref{sec:conclusions}.

\section{Models of quantum-thermalizing machines}\label{sec:model}

Throughout this section we present some properties of the thermalization dynamics associated with {\color{black} two different types of quantum probes embedded in non-thermal baths and} subject to diverse dissipative and decohering processes. \textcolor{black}{In the spirit of the references~\cite{singlemachine,Scarani}, we call our scheme as quantum-thermalizing machines, in which a central system relax to a thermal equilibrium state upon repeatedly interacting with a series of reservoir particles.} Similar results have also obtained by the authors in Ref.~\cite{Angsar}
in the absence of a thermal photon bath and decoherence. These properties
shall be the basis and motivation for further 
study of the power spectrum associated with 
the probe in Sec.~\ref{sec:spectrum} 
and the respective characterization of them by means of spectroscopic measurements.

\subsection{Single qubit probe for an non-thermal bath of $N$-qubit clusters}\label{one_qubit}

First, let us consider our central system as a 
single-qubit, with self-Hamiltonian 
$H_{q}\!=\!\hbar\omega_q\sigma_z/2$
of frequency $\omega_q$, embedded in a thermal 
photon bath and also interacting collectively
du\-ring a time ``$\tau$" at a 
rate $p$
with an environment made of $N$ identical 
qubits [see Fig.~\ref{setup}(a)]
that we will call it {the cluster}
$H_{\rm cl}\!=\!\sum_{j}\hbar\omega_q\sigma_z^j/2$, 
through a dipolar interaction 
$H_{\rm dip}$\!=\!$\sum_{j}\hbar g(\sigma_+^j\sigma_-$+$\sigma_-^j\sigma_+)$.
Here, $g$ is the interaction strength, 
$\sigma_z^j$ and $\sigma_\pm^j$ are the 
usual Pauli $z$ and ladder operators of the  
$j=1,2,..N$-th qubit, respectively with
$[\sigma_z^i,\sigma_\pm^j]\!=\!\pm 2\sigma_\pm^i\delta_{ij}$
and
$[\sigma_+^i,\sigma_-^j]\!=\!\sigma_z^i\delta_{ij}$. Without ambiguity, operators without superscript $j$ will
refer to the central qubit system.
Following the approach adopted for the micromaser 
theory~\cite{Filipowicz}, we will assume that before 
each interaction the state of the 
$N$-qubit cluster has to be reset to its initial
state $\rho_{\rm cl}$.
Under this assumption, it is possible to derive 
a Markovian master equation for the 
single-qubit density matrix $\rho_q$.
Thereby, tracing out the degrees of freedom 
of the cluster \textcolor{black}{and the thermal bath},
in the interaction picture 
associated with the self Hamiltonian of the 
central qubit, and up to second order in 
$g\tau$ we get 
(see Appendix~\ref{apx_micro} for a detailed 
derivation):
\begin{eqnarray}\label{master1}
\dot\rho_q=r_d\mathcal{L}[\sigma_-]\rho_q
+r_e\mathcal{L}[\sigma_+]\rho_q+
{\gamma_\phi}\mathcal{L}[\sigma_z/4]\rho_q,
\end{eqnarray}
where 
$\mathcal{L}[o]\rho_q\equiv 2o\rho_qo^\dagger
-o^\dagger o\rho_q-\rho_qo^\dagger o$ 
is the usual Lindblad super-operator.
$r_e=[\mu_q\langle J_+J_-\rangle+\gamma\bar{n}_{\rm en}]/2$ 
and 
$r_d=[\mu_q\langle J_-J_+\rangle+\gamma(\bar{n}_{\rm en}+1)]/2$ 
are the excitation (heating) and 
de-excitation (cooling) rates, respectively, with
$\mu_q=p(g\tau)^2$ is the effective coupling rate and
$\bar{n}_{\rm en}=(\exp[\hbar\omega_q/(k_BT_{\rm en})]-1)^{-1}$
is the ave\-ra\-ge photon number of a possibly unavoidable background
thermal environment 
at temperature $T_{\rm en}$ that it is in contact
with the central qubit system. 
$J_\pm=\sum_j\sigma_\pm^j$ and $J_z=\sum_j\sigma_z^j/2$ 
are the collective spin operators~\cite{Ban93}.
It is known that dissipation of energy and 
de\-pha\-sing have a dramatic impact on the 
performance of quantum thermal and non-thermal 
machines. For instance, making use of dissipation 
in the photon field  and pure atomic dephasing 
the quantum-classical transition of a photon-Carnot 
engine was evidenced in~\cite{QCtranPRE}. 
Here, those decoherence processes are incorporate,
phe\-no\-me\-no\-logi\-cally in Eq.~\eqref{master1}, 
through the qubit spontaneous emission (SE) coefficient 
$\gamma$ and a nonradiative dephasing rate 
$\gamma_\phi$.

Coherences of $\rho_\text{cl}$ that contribute to 
the average value 
$\langle J_-J_+\rangle={\rm Tr}\{\rho_{\rm cl}J_-J_+\}$ 
and its complex conjugate were identified as the HECs of the $N$-qubit cluster~\cite{Angsar,BarisThermal,CerenEntropy} which, in turn, act as an environment with an apparent temperature to the central qubit system. 
Therefore, along this work, we will be 
interested in determining
$\langle J_\pm J_\mp\rangle$
and their influence on the quantum evolution 
of the central system.

Equation~\eqref{master1} has the following 
solution for its density matrix 
elements~\cite{breuer2002theory}:
\begin{subequations}\label{solmasqubit}
\begin{align}
\rho_{eg}(t)&=\rho_{eg}(0)\exp\left[-\left(r_d+r_e+\gamma_\phi\right)t\right],\label{correl} \\ 
\rho_{ee}(t)&=\frac{[r_d\rho_{ee}(0)-r_e\rho_{gg}(0)]e^{-2t\left(r_d+r_e\right)}+r_e}{r_d+r_e},
\label{popu}
\end{align}
\end{subequations}
$\rho_{ge}(t)=\rho_{eg}(t)^*$ and 
$\rho_{gg}(t)=1-\rho_{ee}(t)$.
We have used the notation
$\rho_{eg}=\langle e|\rho_q|g\rangle$,
$\rho_{ee}=\langle e|\rho_q|e\rangle$,
$\rho_{gg}=\langle g|\rho_q|g\rangle$
and $\rho_{ge}=\langle g|\rho_q|e\rangle$, 
where $|g\rangle$ ($|e\rangle$) is the 
ground (excited) qubit state. 

\textcolor{black}{Defining the internal energy of 
the system as $E_q={\rm Tr}\{\rho_q H_q\}$,
the heat flow from a non-thermal bath to the
qubit probe is given by 
$\dot{Q}_q=\dot{E}_q={\rm Tr}\{\dot\rho_q H_q\}$.
Using Eq.~(\ref{master1}) it is easy to show 
that $\dot{Q}_q=2\hbar\omega_q\big[r_e\rho_{gg}(t)-r_d\rho_{ee}(t)\big]$,
which can be rewritten in the following form
\begin{equation}\label{heat_flow_qubit}
\dot{Q}_q=2\hbar\omega_qr_d\rho_{gg}\big[\exp(-\hbar\omega_q/k_BT_{ B})-\exp(-\hbar\omega_q/k_BT_q)\big],
\end{equation}
where $k_B$ is the Boltzmann constant,
$T_q$ and $T_{B}$ are the apparent
temperatures of the qubit probe and the 
non-thermal bath respectively.
These are defined as~\cite{Latune_2019}
\begin{equation}\label{eq:temp}
T_q=\frac{\hbar\omega_q}{k_B}\Big[\ln\left(\frac{\rho_{gg}(t)}{\rho_{ee}(t)}\right)\Big]^{-1},\quad
T_B=\frac{\hbar\omega_q}{k_B}\Big[\ln\Big(\frac{r_d}{r_e}\Big)\Big]^{-1}.
\end{equation}
One can also see that $T_q$ and $T_B$ can be viewed as temperatures due to the fact that 
they determine the sign of the heat current, i.e. $\dot Q_q$ is positive if 
$T_q\geq T_B$. Notice that $T_q$ ($T_B$) is, in principle, 
a time-dependent (time-independent) parameter that depends on the state of the probe (bath), which can be thermal or non-thermal. In fact, if we switch-off the
repeated interaction with the $N$-qubit 
cluster, $\mu_q=0$, the apparent temperature
$T_B$ is equal to the temperature of the
background thermal environment $T_{\rm en}$.
In the steady state $\rho_q^{st}$ is a statistical 
mixture of its excited and ground states with 
probabilities  
$\rho_{ee}^{st}=\rho_{ee}(\infty)=r_e/(r_d+r_e)$ 
and
$\rho_{gg}^{st}=\rho_{gg}(\infty)=r_d/(r_d+r_e)$
respectively. Moreover, 
$\rho_{gg}^{st}/\rho_{ee}^{st}=r_d/r_e$,
therefore, at the steady state the qubit 
probe thermalise at the apparent temperature 
of the non-thermal bath.}
Hence, \textcolor{black}{the} qubit temperature can be controlled by 
the ratio between the cooling and heating rates. $T_q$ 
\textcolor{black}{can} also be obtained 
\textcolor{black}{only} 
by looking for the population of the qubit 
excited steady state since 
$T_q=(\hbar\omega_q/k_B)[\ln(\rho_{ee}^{st-1}-1)]^{-1}$.
\textcolor{black}{Note that the apparent temperature of the qubit probe can have negative values at the steady state, i.e. it is possible to invert the populations of the probe, if $r_e>r_d$. This condition translates into the expectation values in the following way $\langle J_+J_-\rangle -\langle J_-J_+\rangle>\gamma/\mu_q$ and is only satisfiable if we  have an inverted population in the bath states. Eventhough HECs have no role in determining this condition since they contribute to both expectation values with the same amount, they can be used as a knob to enhance the negative apparent temperature through their dependence on the apparent temperature $r_d/r_e$, as presented in Eq.~(\ref{eq:temp}). Significance of this result is that even a set of weakly inverted qubits, slightly above the effective infinite temperature, can be used as a fully inverted system, with the aid of sufficient quantum coherence.}

From Eqs.~\eqref{correl} and \eqref{popu}, we see
that $\gamma_\phi$ only affects the off-diagonal 
density matrix elements, which implies that 
dephasing does not change either $T_q$ or the 
qubit thermalization time  $t_q\equiv[2(r_d+r_e)]^{-1}$.
The latter can be rewritten as
\begin{align}\label{qubittime}
t_q=\frac{1}{\gamma(2\bar{n}_{\rm en}+1)+2\mu_q(\langle J^2\rangle-\langle J_z^2\rangle)},
\end{align}
where $J^2=J_z^2+(J_+J_-+J_-J_+)/2$ is the
{\em Casimir} operator of 
$\mathfrak{su}(2)$~\cite{Ban93}.

\subsection{Single cavity probe for an non-thermal bath of $N$-qubit clusters}\label{one_cavity}

Now we replace the single-qubit central system 
from the pre\-vious example by a single 
electromagnetic field mode 
$H_c=\hbar\omega_c a^\dagger a$ 
of frequency $\omega_c$ inside a leaky cavity.
The resonant collective interaction 
(of strength $\lambda$) between the field and 
the $N$-qubit coherent clusters,
is governed by the Tavis-Cummings (TC) Hamiltonian
$H_\text{TC}=\lambda(J_+a+J_-a^\dagger)$~\cite{Tavis}.
Here, $a$ ($a^\dagger$) is the usual annihilation
(creation) bosonic field ope\-rator that satisfy 
$[a,a^\dagger]=1$. 
\textcolor{black}{This kind of collective 
interaction has been realized experimentally 
with $N=3$~\cite{Fink} (and $N=6$~\cite{arXiv_Yang})
fully controllable superconducting qubits embedded 
in a transmission line resonator, where the coupling
strengths between the qubits and the resonator were 
virtually identical.}
A schematic representation of the TC model is 
sketched in Fig.~\ref{setup}(b).
Under the standard assumptions of the micromaser
theory~\cite{Filipowicz,scully_zubairy,Temnov} and 
within the aforementioned conditions, the 
following Markovian master equation for the
electromagnetic field density matrix $\rho_c$ 
(in a rotating frame at the cavity frequency) 
can be derived~\cite{CerenEntropy,Esposito_PRX_2017} 
(see Appendix~\ref{apx_micro}):
\begin{align}\label{mastercavity}
\dot{\rho}_c=\mathcal{R}_d\mathcal{L}[a]\rho_c+
\mathcal{R}_e\mathcal{L}[a^\dagger]\rho_c
+\kappa_\phi\mathcal{L}[a^\dagger a]\rho_c,
\end{align}
Once again,
$\mathcal{R}_d$=$[\mu_c\langle J_-J_+\rangle+\kappa(\bar{n}_{\rm en}+1)]/2$
and $\mathcal{R}_e$=$[\mu_c\langle J_+J_-\rangle+\kappa\bar{n}_{\rm en}]/2$
represent the cooling and heating rates 
respectively. 
$\kappa$ ($\kappa_\phi$) is the cavity field
decay (dephasing) rate and $\bar{n}_\text{en}$
now depends on the $\omega_c$ instead of 
$\omega_q$. Like in the previous case 
{\color{black} $\mu_c=p(\lambda\tau)^2$} is the 
effective coupling rate.
To elucidate how the HECs intervene in the 
cavi\-ty field evolution, we first calculate 
the co\-rres\-ponding equations of motion for 
the photon number $a^\dagger a$ and field $a$ 
operators, these are given by
$\langle \dot{a^\dagger a}\rangle_t =-2(\mathcal{R}_d-\mathcal{R}_e)
\langle a^\dagger a\rangle_t+2\mathcal{R}_e$
and
$\langle \dot{a}\rangle_t =-(\mathcal{R}_d-\mathcal{R}_e
+\kappa_\phi)\langle a\rangle_t$.
The notation 
$\langle \mathcal{O}\rangle_t=
{\rm Tr}\{\rho_c \mathcal{O}(t)\}$ was used.
Taking the initial conditions 
$\langle a^\dagger a\rangle_0$ and 
$\langle a\rangle_0$ their solutions are:
$\langle a^\dagger a\rangle_t =
n_{st}+(\langle a^\dagger a\rangle_0-n_{st})
\exp[-2(\mathcal{R}_d-\mathcal{R}_e)t]$ 
and
$\langle a\rangle_t =\langle a\rangle_0 
\exp[-(\mathcal{R}_d-\mathcal{R}_e+\kappa_\phi)t]$,
where
$n_{st}=\mathcal{R}_e(\mathcal{R}_d-\mathcal{R}_e)^{-1}$
is the steady state photon number.
Substituting the values of the excitation 
and de-excitation rates yields
$n_{st}$=$({\kappa\bar{n}_{\rm en}+\mu_c\langle J_+J_-\rangle})$/
$({\kappa -2\mu_c\langle J_z\rangle})$.
The heat flow from a non-thermal
bath to the field mode is 
$\dot{Q}_c={\rm Tr}\{\dot{\rho}_cH_c\}$. 
Using Eq.~(\ref{mastercavity}) this reduces to
$\dot{Q}_c=2\hbar\omega_c\big(\mathcal{R}_e\langle aa^\dagger\rangle_t-
\mathcal{R}_d\langle a^\dagger a\rangle_t\big)$.
Rewriting this expression in the form of 
Eq.~(\ref{heat_flow_qubit}), we can obtain
the apparent temperature of the field mode
given by 
$T_c=(\hbar\omega_c/k_B)[\ln\big(\langle a^\dagger a\rangle_t^{-1}+1\big)]^{-1}$.
As expected, at the steady state, the field
probe thermalise at the apparent temperature 
of the non-thermal bath, i.e., 
$T_B=(\hbar\omega_c/k_B)\ln(\mathcal{R}_d/\mathcal{R}_e)^{-1}$.
On the other hand, thermalization time of the 
field mode is given by
$t_c\equiv [2(\mathcal{R}_d-\mathcal{R}_e)]^{-1}$,
which can be rewritten in terms of the total 
popu\-lation inversion $\langle J_z\rangle$ as:
\begin{align}\label{cavitytime}
t_c=\frac{1}{\kappa-2\mu_c\langle J_z\rangle}.
\end{align}
The interaction of the central system with the
$N$-qubit coherent cluster can modify its natural 
(or fabricated) parameters. 
For instance, if 
$\kappa_\phi=0$, the coherence time of 
field mode is equal to $2t_c$, thus, 
from Eq.~\eqref{cavitytime} it is clear that $2t_c$
can be enhanced just by having, for example, a 
large number of environmental qubits in their 
excited state, this could improve 
the performance of a thermal machine using the
field mode as the probe. 
Such a situation is especially useful when 
changing the quality factor of a cavity 
(or transmission line resonator) is difficult 
due to manufacturing limitations.

Lastly, we would like to comment on the possibility of negative temperatures in the present case. Such a situation only occurs when $\mathcal{R}_d<\mathcal{R}_e$, however this condition implies that the thermalization time $t_c$ can be infinite or even negative. Moreover, the steady-state photon number $n_{st}=\mathcal{R}_e(\mathcal{R}_d-\mathcal{R}_e)^{-1}$ also beomes ill-defined. These problems have their roots in the fact that it is not possible for a quantum harmonic oscillator to attain negative temperatures, since its energy spectrum is not bounded from above.

\subsection{Bath size effects on the probe thermalization}

So far, we can point out some important things 
regarding thermalization of the single-qubit 
system and the single-field mode. 
Despite the fact that $T_q$ and $T_c$ look different, 
they are defined as the logarithm of the 
ratio between their respective 
{\em heating} and {\em cooling} rates.
Consequently, both tempera\-tures have the same 
depen\-dency on $\langle J_\pm J_\mp\rangle$.
This implies that the only difference between 
both temperatures should be attri\-bu\-ted
to the corresponding parameters that characterize
each physical system.
However, there must be some situ\-ations, of 
practical purposes, in which it would be more 
advantageous to work with a cavity field mode 
(resonator) instead of a single-qubit as the 
main central system or vice versa.
As an example, let's consider the initial 
state of $N$-qubit cluster to be a Dicke state 
$\rho_\text{cl}=|k,N\rangle\langle k,N|$. This
is a non-thermal state prepared by non-thermal 
means where $k$ is the number 
of excitations in the cluster and 
$\langle J_+J_-\rangle=k(N-k+1)$,
$\langle J_-J_+\rangle=(k+1)(N-k)$ and
$\langle J_\pm\rangle=\langle J_\pm^2\rangle=0$~\cite{klimov2009group}. 
If one considers a Dicke state that is close to the central block, i.e. setting $k=N/2-1$, 
it is possible to see from the above expressions that $T_q$ and $T_c$ grow 
with $N^2$, see Ref.~\cite{Angsar} for a detailed derivation. This superlinear scaling is fundamentally based upon symmetries of Dicke-type states. An easy interpretation could be their relation to possible ways of selecting pairs of states among N degenerate ones, which scales as $N^2$. However, the corresponding thermalization times (using the Dicke states and $k=N/2-1$) are
quite different; $t_q$ is inversely proportional to 
$N^2$~\cite{Angsar} and $t_c$ will be independent of the number
of environmental qubits.
Hence, if there is a difference between the
thermalization time of a bosonic and fermionic 
probes, this should be relevant for the 
design of future quantum thermal and non-thermal
machines. For instance, 
any shortening in $t_q$ or $t_c$ could be used 
to boost the performance of a quantum heat engine operating at a 
finite time~\cite{KoslDyna}.
Therefore, it is necessary to know how to
check this behavior experimentally, in Sec.~\ref{sec:spectrum} we will
address this question from the spectroscopic 
point of view.

\subsection{Case study: $N\!=\!2$ qubit cluster}\label{sec:casestudy}

\begin{figure}[tp]
\includegraphics[width=7.5cm, height=7.5cm]{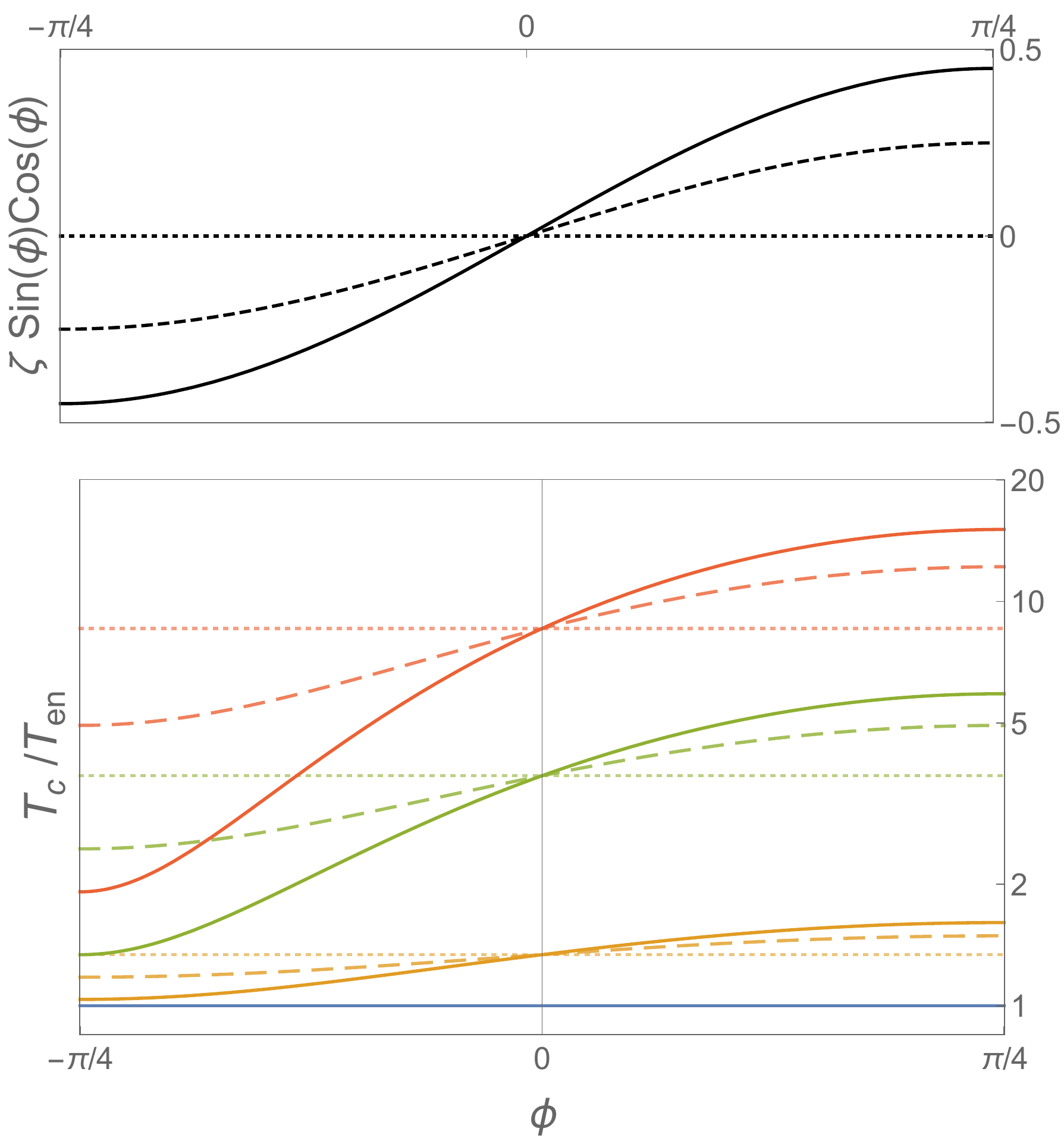} 
\caption{(Color online) Above: HEC of 
the state~\eqref{clus_state} as a function of $\phi$ 
for purity parameter $\zeta= \{0.9, 0.5, 0\}$ 
(solid, dashed, dotted).
Below: ratio between cavity field temperature $T_c$ 
and the temperature of its
local environment at $T_\text{en}$. Relative 
interaction strength $\mu_c/\kappa$ 
=$\{0,0.1,1,3\}$ (blue, orange, green, red). 
Average thermal photon number is $\bar{n}_\text{en}=0.1$, 
which could correspond to $T_\text{en}\sim$ mK 
($\sim 298$~K) at microwave (optical) frequencies.\hspace{6.1cm}~}
\label{concu}
\end{figure}

The results obtained in the previous section are quite general.
In order to clearly demonstrate these results, which shows a transpa\-rent link between the HECs and these quantum 
thermalization effects, we now choose a specific size for the cohe\-rent cluster, $N=2$, with the 
following initial state (written in the standard 
two-qubit basis $[ee,eg,ge,gg]$) 
\begin{align}\label{clus_state}
\rho_\text{cl}({\phi,\zeta})=
\begin{pmatrix}
0&0&0&0\\
0&\sin^2\phi&\zeta\sin\phi\cos\phi&0&\\
0&\zeta^*\sin\phi\cos\phi&\cos^2\phi&0\\
0&0&0&0
\end{pmatrix},
\end{align}
where $\zeta$ is a kind of purity parameter such 
that $|\zeta|\leq 1$ and $\phi\in[-\pi/4,\pi/4]$.
For $\zeta$=$0$ ($\zeta=1$) \revision{the} Eq.~\eqref{clus_state}
reduce to a statistical mixture (Bell-like state). 
This state will help us to make the connection between the HECs and the properties of the power spectrum of the probe system in the next section, Sec.~\ref{hec_spec}. 
Accor\-dingly, $\rho_\text{cl}({\phi,\zeta})$ gives the following result for the expectation values
\begin{align}\label{hec_rho}
\langle J_\pm J_\mp\rangle=1+2 {\color{black}{\rm Re}(\zeta)} \sin\phi\cos\phi,
\end{align}
and
$\langle J_\pm\rangle\!=\!\langle J_\pm^2\rangle\!=\!0$.
\textcolor{black}{Without loss of genera\-lity, we will assume $\zeta$ to be a non-negative real number.}
We identify the off-diagonal terms $\zeta\sin\phi\cos\phi$ 
of $\rho_\text{cl}({\phi,\zeta})$ as the so 
called HECs since they contribute to the 
{\em heat\-ing} and {\em cooling} rates through 
Eq.~\eqref{hec_rho}.
First, we proceed to examine their action on the 
probe temperature.
From the discussion in Sec.s~\ref{one_qubit} and \ref{one_cavity}, we know that the thermalization temperatures of both qubit ($T_q$) and cavity ($T_c$) probe are the same. Therefore, we decided to exemplify the HEC dependence of final temperature by working with a cavity probe.  
Using Eq.~\eqref{clus_state}, we find $\langle J_z\rangle=0$
and the steady state photon number reads as
$n_{st}=\bar{n}_\text{en}+({\mu_c}/{\kappa})(1+\zeta\sin 2\phi)$.
Thus, any change in $n_{st}$ due to $\zeta$ or
$\phi$ will be reflected directly in $T_c$.
This can be seen in Fig.~\ref{concu}, where we 
show how the ratio of the final temperature of the cavity to the thermal environment temperature, $T_c/T_{\rm en}$, and HECs, change as a function of 
$\phi$ for different values of $\zeta$ and 
relative interaction strength $\mu_c/\kappa$. In fact,
Fig.~\ref{concu} demonstrates how $T_c$ differs,
as expected~\cite{CerenEntropy}, when the 
interaction with the $N$-qubit coherent cluster is on as compared to the case where only thermal bath and incoherent clusters are present (dotted lines). In addition, depending on the sign of coherences of $\rho_{\rm cl}(\phi,\zeta)$ it is possible to heat (cool) the probe above (below) the temperature of the thermal bath with maximum attained when $\phi=\pi/4$ ($-\pi/4$) for a given value of $\zeta$. 

In contrast to final temperatures, thermalization 
times are distinct when again
$\rho_{\rm cl}(\phi,\zeta)$ is considered as the non-thermal bath.
Specifically, Eq.~\eqref{cavitytime} redu\-ces to 
$t_c^{-1}=\kappa$ and Eq.~\eqref{qubittime} to
$t_q^{-1}=\gamma(1+2\bar{n}_\text{en})+2\mu_q(1+\zeta\sin 2\phi)$,
due to $\langle J^2\rangle-\langle J_z^2\rangle=1+\zeta\sin 2\phi$
for the latter case. 

A comment on the possibility of negative temperatures is in order. The previously mentioned condition for negative apparent temperatures, $r_e>r_d$ takes the form $2\mu_q(\rho_{\uparrow\uparrow}-\rho_{\downarrow\downarrow})>\gamma$ in the case of two-qubit clusters, where $\rho_{\uparrow\uparrow}$ and $\rho_{\downarrow\downarrow}$ are both excited and ground state populations of qubits, respectively. From this inequality we can conclude that in order to achieve a population inversion of the probe, one needs to have a higher population in the excited states of the cluster than it has in the ground state and this difference needs to be large enough to overcome the effect of the thermal bath characterized by the spontaneous emission rate $\gamma$. Clearly, the specific cluster state we consider in Eq.~(\ref{clus_state}) does not satisfy this condition therefore is unable to cause a population inversion, but it is possible to achieve negative probe temperatures with different initial cluster states. Naturally, for larger clusters, the above condition will be modified, however, the conclusion that we need higher populations in which the majority of the cluster is in excited state, still stands.

Before moving on to the next section, we would like make a small comment on the correlation and coherence properties of considered coherent cluster state. The entanglement content of Eq.~\eqref{clus_state}, as quantified by the concurrence~\cite{Wootters}, can be found as $\mathcal{C}=2\zeta|\sin\phi\cos\phi|$. In the present case, the entanglement of the cluster turns out to be equal to the $l_1$-norm of coherence~\cite{PlenioCohe}, which is defined as the sum of absolute values of all off-diagonal elements of the state. Therefore, it is possible to conclude that these quantities actually have a direct effect in the final temperatures of the probes. Although broad temperature control in cavi\-ties by combustion of two-atom entanglement has already been predicted in~\cite{gerzontemp}; in this work, {\color{black} we have presented an example of those results for the case in which the quantum probe has a $\mathfrak{su}(2)$ symmetry. We would like to emphasize once again that the coincidence of entanglement and $l_1$-norm only applies to the presented example. In general, such a relation may not hold true. Moreover}, we have showed that these different cases yield completely different thermalization times for the same amount of coherences in the non-thermal environment. 

\section{Spectral measurements}\label{sec:spectrum}

Up to this point, we have verified that changes 
in the parameters controlling the HECs or in the size of
the cluster have a direct impact on the 
temperature and thermalization time of the 
corresponding probe. Now, we present a way to experimentally test such predictions.
In this section, we will address such an inquiry 
using the power spectrum of the probe.

\subsection*{Stationary power spectrum}

The stationary (Wiener-Khintchine) power 
spectrum~\cite{wiener1930,Kinchin} associated 
with the single-qubit probe, 
is defined as~\citep{carmichael,castro2016}:
\begin{align}\label{specdef1}
S_q(\omega)=\frac{1}{\pi} {\rm Re}\int_0^\infty d\tau
e^{i\omega\tau}\langle \sigma_+(0)\sigma_-(\tau)\rangle,
\end{align}
where $\langle \sigma_+(0)\sigma_-(\tau)\rangle\!=\!\lim_{t\rightarrow\infty}\langle \sigma_+(t)\sigma_-(t+\tau)\rangle$
is the first order dipole-field auto-correlation 
function. In the steady state of
the system at hand, it can be calculated as (see Appendix~\ref{apx_corr}):
$\langle \sigma_+(0)\sigma_-(\tau)\rangle\!=\!\rho_{ee}^{st}\exp[-(i\omega_q+\gamma_\phi+(2t_q)^{-1})\tau]$.
Using this result, we can perform the integral in 
Eq.~\eqref{specdef1} and obtain the stationary power spectrum of the qubit as follows
\begin{align}\label{qub_spec}
S_q(\omega)=\frac{\rho_{ee}^{st}}{\pi}\frac{\frac{1}{2}\Gamma_q}
{\big(\omega-\omega_q\big)^2+\big(\frac{1}{2}\Gamma_q\big)^2},
\end{align}
which is a Lorentzian peak of full width 
at half maximum (FWHM) given by
$\Gamma_q=2\gamma_\phi+t_q^{-1}$. Further, we can use Eq.~\eqref{qubittime} to replace the thermalization time and obtain
\begin{align}\label{qub_band}
\Gamma_q=2\gamma_\phi+\gamma(2\bar{n}_{\rm en}+1)+2\mu_q(\langle J^2\rangle-\langle J_z^2\rangle).
\end{align}
$S_q(\omega)$ has a maximum at $\omega_q$ given 
by $2\rho_{ee}^{st}/(\Gamma_q\pi)$.
The intensity can be calculated as the integral over all frequencies
$I_q\equiv\int_{-\infty}^\infty S_q(\omega)d\omega=\rho_{ee}^{st}$. \textcolor{black}{It is important to note that, since it is equal to the excited state population, any value of the intensity that is greater than $1/2$ indicates that the probe temperature has attained negative values.}

On the other hand, the power spectrum of the 
single-field mode is defined as
\begin{equation}\label{specdef2}
S_c(\omega)=\frac{1}{\pi}{\rm Re}
\int_0^\infty d\tau e^{i\omega\tau}
\langle a^\dagger(0)a(\tau)\rangle,
\end{equation}
and if we repeat the same procedure as we did for the qubit case, it is possible to find the power spectrum as 
\begin{align}\label{cav_spec}
S_c(\omega)=\frac{n_{st}}{\pi}\frac{\frac{1}{2}\Gamma_c}
{\big(\omega-\omega_c\big)^2+\big(\frac{1}{2}\Gamma_c\big)^2},
\end{align}
with its maximum and spectral bandwidth being $S_c(\omega_c)=2n_{st}/(\Gamma_c\pi)$ and $\Gamma_c=2\kappa_\phi+t_c^{-1}$, respectively. Again, similar to the qubit case, we can substitute the thermalization time of the cavity from Eq.~\eqref{cavitytime} to obtain
\begin{align}\label{cav_band}
\Gamma_c=2\kappa_\phi+\kappa-2\mu_c\langle J_z\rangle,
\end{align}
resulting in a cavity intensity of $I_c\equiv\int_{-\infty}^\infty S_c(\omega)d\omega=n_{st}$.
We observe that in both cases the spectral bandwidth depends explicitly on the corresponding thermalization
times. Hence, by measuring the FWHM of these spectral signals, it is possible to infer how fast or 
slow is the thermalization process. Moreover, measuring 
their intensity (signal integration) one can 
obtain the apparent temperature
that has been reached by the quantum probe.

If we rewrite the ratio $r_d/r_e$ 
($\mathcal{R}_d/\mathcal{R}_e$) and the sum 
(difference) $r_d+r_e$ 
($\mathcal{R}_d-\mathcal{R}_e$) in terms of  
$I_q$ ($I_c$) and $\Gamma_q$ ($\Gamma_c$)
respectively, we can arrange simple and general
expressions for the average values we are 
interested in:
\begin{subequations}\label{aver_values}
\begin{align}
\langle J_+J_-\rangle &=(\Gamma_\text{p} -2\gamma^\phi_\text{p})({I}_\text{p})\mu_\text{p}^{-1}
-\frac{\gamma_\text{p}}{\mu_\text{p}}\bar{n}_{\rm en},\label{aver_values_b}  \\  
\langle J_-J_+\rangle &=(\Gamma_\text{p} -2\gamma_\text{p}^\phi)
(1\pm{I}_\text{p})\mu_\text{p}^{-1}
-\frac{\gamma_\text{p}}{\mu_\text{p}}(1+\bar{n}_{\rm en}),\label{aver_values_b2}
\end{align}	
\end{subequations}
where the subscript ``$\text{p}$'' specify the type of the probe,  i.e., $q$ or $c$. 
In the same manner, decay rate 
$\gamma_\text{p}$ 
($\gamma_\text{p}^\phi$) must be replaced 
by $\gamma$ or $\kappa$ 
($\gamma_\phi$ or $\kappa_\phi$).
The change of sign in Eq.~\eqref{aver_values_b} 
comes from considering the single-qubit instead 
of the cavity mode as the central system.
Eqs.~\eqref{aver_values} are useful because they enable, together with some previous knowledge of the bath state like Eq.~(\ref{hec_rho}), to establish a link between spectroscopic parameters, $\Gamma_\text{p}$ and $I_\text{p}$ and HECs, therefore provide the information on whether or not HECs are present in the environment that our probe system is interacting. As the collective spin correlators carry information on the multipartite coherences and entanglement among the environment qubits, our results could allow for spectroscopic quantum thermometry of quantum correlations in a non-thermal environment that can be associated with an apparent temperature. In the following subsection, we will elaborate on these results.

\subsection*{Heat exange coherences in the power spectrum}\label{hec_spec}

\begin{figure}[b]
\includegraphics[width=7.8cm, height=5.5cm]{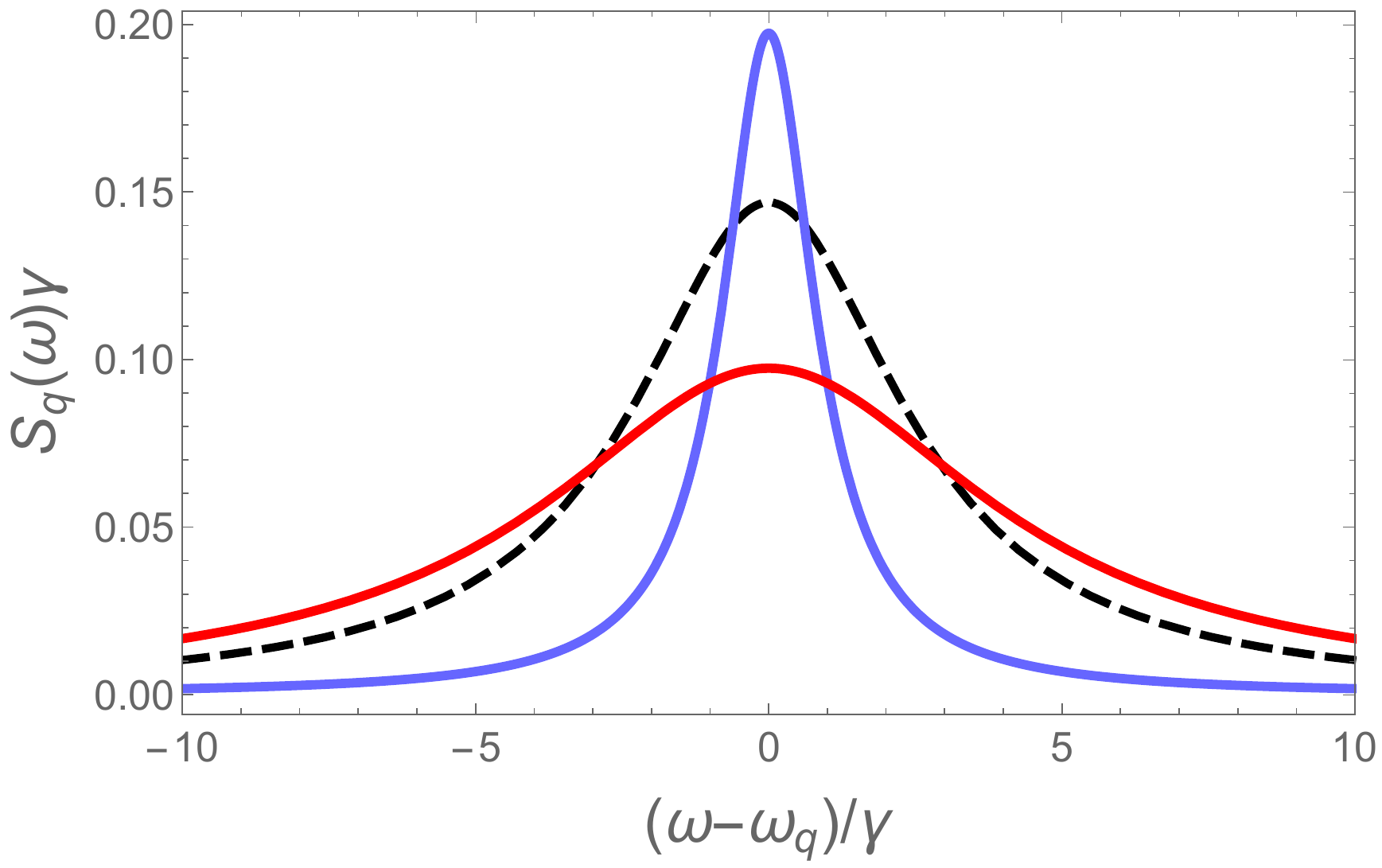} 
\caption{(Color online) Power spectrum of the 
single-qubit probe 
[see Eq.~\eqref{qub_spec}] when the initial state
of the environmental cluster is the 
Eq.~\eqref{clus_state}.
Black dashed line is for $\zeta=0$ (mix state). 
Solid lines are for $\zeta=0.9$ (state with HECs)
and $\phi=\{-\pi/4,\pi/4\}$ (blue, red) represent,
respectively, the coldest and hottest effective
qubit temperature for this value of $\zeta$.
We set $\mu_q=2\gamma$, $\gamma_\phi=0.15\gamma$
and $\bar{n}_{\rm en}=0.1$. The spectral 
bandwidth and the amplitude have a strong 
dependency on $\zeta$ and $\phi$.\hspace{7.6cm}~} 
\label{spec0}
\end{figure}

Now we use the spectra of Eqs.~\eqref{qub_spec} 
and~\eqref{cav_spec} to show how it is possible 
to differentiate between states of the $N$-qubit 
cluster having HECs from those who are a 
statistical mixture. For clarity we consider 
again the state described by 
Eq.~\eqref{clus_state} as an example, from 
Eq.~\eqref{qub_band} follows
\begin{align}\label{gama_qubit_example}
\Gamma_q=2\gamma_\phi +\gamma(2\bar{n}_\text{en}+1)
+2\mu_q(1+\zeta\sin 2\phi).
\end{align}
The inclusion of HECs in Eq.~\eqref{clus_state} 
will (depending on the sign of $\sin 2\phi$) 
increase or decrease the spectral bandwidth 
by an amount
$\Delta \Gamma_q\equiv\Gamma_q-\Gamma_q'=
2\zeta\mu_q\sin2\phi$, where $\Gamma_q'$ is
the FWHM of the mix state $\rho_{\rm cl}(\phi,0)$.
The increase (decrease) is maximum when 
$\zeta=1$ and $\phi=\pi/4$ ($-\pi/4$). 
For $\phi=-\pi/4$ the bandwidth can even be 
equal to the natural (or fabricated) spectral 
linewidth of the qubit as long as $\gamma_\phi$ 
and $\bar{n}_{\rm en}$ are negligible.
This behavior is shown in Fig.~\ref{spec0}
where amplitude variations also can be
observed. 
For the field mode case we cannot apply previous
spectral bandwidth analysis when using 
Eq.~\eqref{clus_state} because 
$\langle J_z\rangle=0$  and  
$\Gamma_c=2\kappa_\phi+\kappa$ remains as if 
there had not been an interaction with the 
$N$-qubit cluster. Howe\-ver, signatures of HECs 
in $S_c(\omega)$ still can be found, these are
encoded in the field intensity given by
$I_c=\bar{n}_{\rm en}+(\mu_c/\kappa)(1+\zeta\sin 2\phi)$. Note that in both cases, HECs affect the same spectral parameters that the temperature of the thermal environment affects. This is in line with our claims that HECs only contribute to the heat flow between the interacting quantum probe and non-thermal environment but cannot be transferred as coherence. All in all, it is nice to see this verification in experimentally accesible spectroscopic parameters.

On the other hand, if the cluster is initially in
the Dicke state $|k,N\rangle$ with $k=N/2-1$, then 
the bandwidth of the single-qubit central system 
depends quadratically on $N$:
\begin{eqnarray}
\Gamma_\text{Dicke}=2\gamma_\phi+\gamma(2\bar{n}_{\rm en}+1)+\mu_q(N+N^2/2-2).
\end{eqnarray}
Thus, in this case, given a set of values
$\{\gamma_\phi,\gamma,\bar{n}_{\rm en}\}$ 
there should be a critical number of qubits in 
the cluster in order to be more significant than
contributions of the dissipation and/or dephasing 
rate. In other words, the control of the spectral
bandwidth is imposed by changes of the cluster size.

These results indicate that, with some previous information of the bath state like Eq.~(\ref{hec_rho}), we are 
capable of discriminating between a 
quantum non-thermal machine operating with a bath containing HECs, from a 
thermal machine in which its environment is in a mixed state.
Even more, we can know if the associated probe temperature is increasing or decreasing together with its rate, just by looking at the spectral properties of it. In fact, presented approach is quite similar to the one put forward in~\cite{arXiv_Paris}, in which the authors suggest to assess the temperature of an environment by looking at the dephasing dynamics of a single, probe qubit that is in contact with that environment. In other words, even though the type of the environment and the interaction with the probe qubit is quite different in the model we consider as compared to~\cite{arXiv_Paris}, the power spectrum of the probe can be used in the same spirit, that is, to estimate the apparent temperature of its environment. However, in this work, the environment that the probe qubit is in contact with do not need to be a thermal one, it can also be a non-thermal, coherent one that dictates an apparent temperature to the probe, through HECs.

Experimental measurements of previous results will be limited by the resolution of the power spectrum~\cite{Clerk_RMP_2010,Caves_PRL_2011} and hence errors in the FWHM and in the intensity would propagate to the errors in coherence and in the apparent temperature respectively. As the errors in power spectrum measurements may arise from technical/systematic errors as well, we ask, more fundamentally, general bounds on estimating the apparent temperature and the HECs of the non-thermal bath by conventional quantum thermometry~\cite{Stace_PRA_2010,dePasquale,arXiv_Sanpera} and estimation theory~\cite{Jing_2014}.

{\color{black} To determine the ultimate limit for which
the apparent temperature $T_p$ of our quantum
probes (qubit or cavity) can be estimated, 
we need to resort to the quantum Cram\'er-Rao bound which reads as~\cite{arXiv_Sanpera}
$\Delta T_p \geq 1/\sqrt{\nu\mathcal{F}(\rho_p)}$,
where $\nu$ is the number of independent 
measurements and $\mathcal{F}(\rho_{p})$ 
is the quantum Fisher 
information (QFI) of the state $\rho_{p}$, 
the latter depends on the parameter $T_p$ that we
want to estimate. When the state of the probe
is a thermal state the QFI can be written in 
terms of the variance of the probe Hamiltonian as~\cite{PRL_Correa} 
$\mathcal{F}(\rho_{p})=(\Delta H_p)^2/(k_B^2T_p^4)$.
Precisely, the steady state solutions of both
Eq.~(\ref{master1}) and Eq.~(\ref{mastercavity})
are thermal states $\rho_p^{st}=\exp(-H_p/k_BT_p)Z^{-1}$
with $Z$ being the corresponding partition function.
Following Ref.~\cite{QST_Steve}, QFI of the qubit 
probe and the field probe (harmonic oscillator),
at the steady state, are given by
$\mathcal{F}(\rho_q^{st})=\big({\hbar\omega_q}/{2k_BT_q^2}\big)^2
{\rm sech}^2\big({\hbar\omega_q}/{2k_BT_q}\big)$
and
$\mathcal{F}(\rho_c^{st})=\big({\hbar\omega_c}/{2k_BT_c^2}\big)^2
{\rm csch}^2\big({\hbar\omega_c}/{2k_BT_c}\big)$
respectively. We present the behaviors of these quantities in
Fig.~\ref{Fisher_probes}.

It is possible to conclude that, for a fixed energy 
gap of the probe, both probes estimate a specific apparent 
temperature with a higher precision than other temperatures.
That specific value of the apparent temperature 
corresponding to the maximum of the QFI is linearly
related with the energy level splitting of the probe,
this is
$2k_BT_p^{\rm max}=\alpha_p\hbar\omega_p$, where
$\alpha_p$ satisfies the transcendental 
equations~\cite{QST_Steve}:
$2\alpha_q=\tanh(\alpha_q^{-1})$ for the 
qubit probe and 
$2\alpha_c=\coth(\alpha_c^{-1})$ for 
the cavity probe. 
These two transcendental equations have 
an approximate solution $\alpha_p\approx 0.5$ ($\alpha_q=0.484$ and $\alpha_c=0.522$ exactly) that
can be inferred, roughly, from their
behaviour in Fig.~\ref{Fisher_probes}.
Thus, the apparent temperature which can be estimated with the maximum precision
(the minimum uncertainty $\Delta T_p$) is 
$T_p^{\rm max}\sim\hbar\omega_p/4k_B$.
\begin{center}
\begin{figure}[t!]
\includegraphics[width=8.0cm, height=5.2cm]{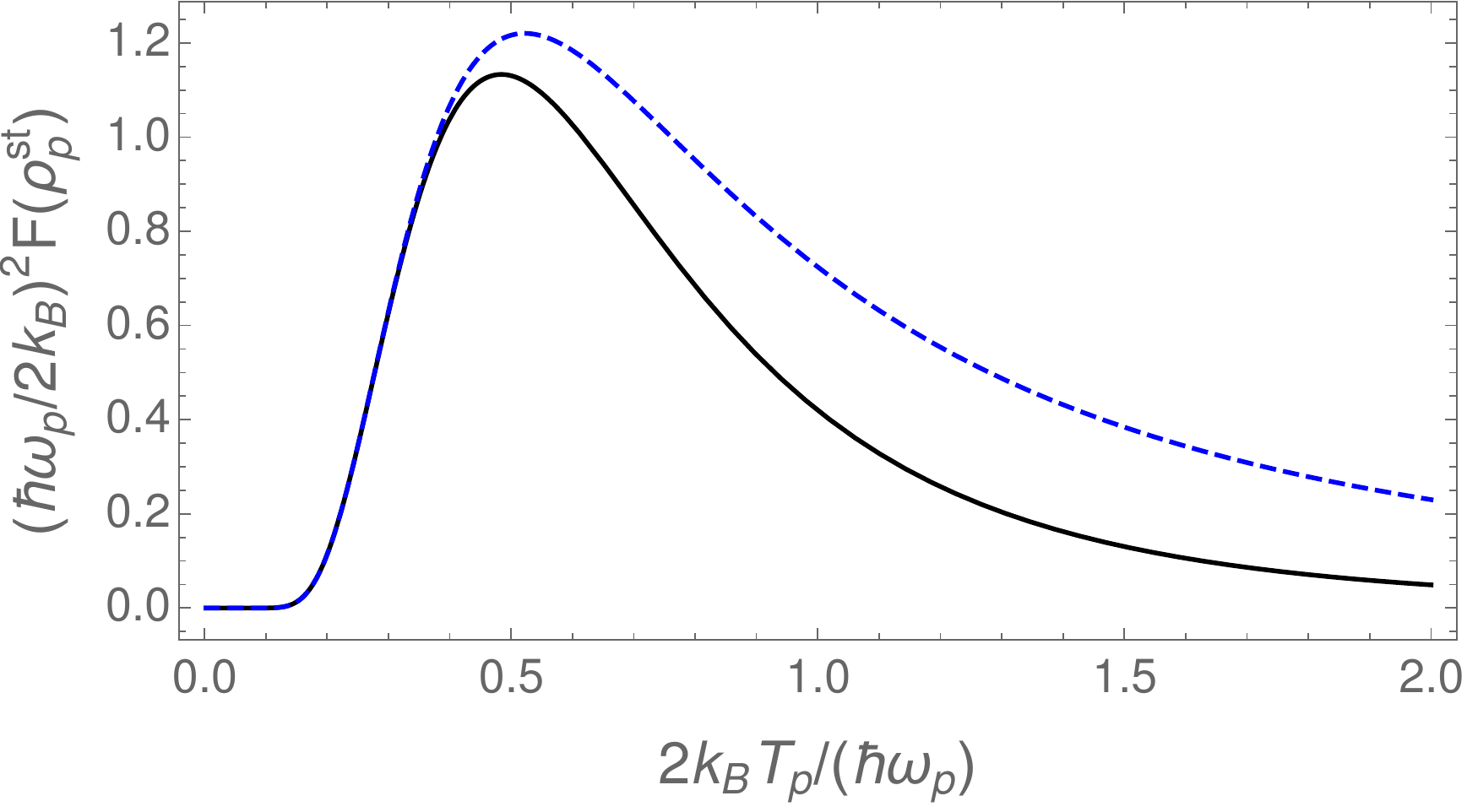} 
\caption{ \textcolor{black}{Quantum Fisher information of each quantum probe. 
Solid-black (blue-dashed) line is for the qubit (ca\-vi\-ty) probe.\hspace{7.5cm}~}}
\label{Fisher_probes}
\end{figure}
\end{center}
In the present scheme, since the apparent temperature is inherently dependent on the value of HECs in the bath, it is also possible to extend the usage of the above information and determine how precisely one can identify the amount of HECs in the bath state.
However, to obtain the uncertainty of the HECs from the uncertainty on the apparent temperature, a proper error propagation treatment should be done. Instead, we reproduce the parameter estimation framework of~\cite{arXiv_Sanpera} and~\cite{Jing_2014} but focusing directly on $\zeta$ (the purity or amplitude parameter in the HECs of the state~(\ref{clus_state})) rather than the apparent temperature. Since $T_p=T_p(\zeta)$, the probabilities  $p_n\equiv\langle \epsilon_n|\rho_p^{st}|\epsilon_n\rangle$ which depend on $T_p$ will also depend on $\zeta$, i.e., $p_n=p_n(T_p(\zeta))$, here $|\epsilon_n\rangle$  is an eigenstate of $H_p$. Following very closely \cite{arXiv_Sanpera,Jing_2014} and making use of the chain rule $\partial p_n/\partial\zeta=(\partial p_n/\partial T_p)(\partial T_p/\partial\zeta)$, it is possible to get a compact and simple form of the QFI associated with the estimation of $\zeta$, in terms of the QFI for thermal states of the probe times the rate of change of the apparent temperature with respecto to $\zeta$:
\begin{equation}\label{qfi_zeta}
\mathcal{F}(\rho_\zeta)=\mathcal{F}(\rho_p^{st})(\partial T_p/\partial\zeta)^2.
\end{equation}
Due to the general form of the above expression, it can be used to estimate any other parameter that is linked with the apparent temperature, such as the phase of the HECs for instance. Using the expression of the apparent temperature for the qubit probe the Eq.~(\ref{qfi_zeta}) can be written, in a more explicit form, in terms of the heating and cooling rates
\begin{equation}
\mathcal{F}(\rho_\zeta)=\frac{r_e}{r_d}\left[\left(\frac{r_e}{r_e+r_d}\right)\frac{\partial}{\partial \zeta}\left( \frac{r_d}{r_e}\right)\right]^2.
\end{equation}
Therefore, the uncertainty in the estimation of $\zeta$ will be bounded by $\Delta\zeta\geq 1/\sqrt{\nu\mathcal{F}(\rho_\zeta)}$.
One can take a step further and give an explicit expression for the $\mathcal{F}(\rho_\zeta)$  for the state given in Eq.~(\ref{clus_state}) as follows
\begin{equation}
\mathcal{F}(\rho_\zeta)=\frac{(\mu_q/\gamma)^2\sin^2(2\phi)}{z(1+z)(1+2z)^2},
\end{equation}
where $z=\bar{n}_{\rm en}+(\mu_q/\gamma)(1+\zeta\sin 2\phi)$.

\section{Displacement coherences in the power spectrum}\label{dis_cohe}

We now want to shift our attention from HECs to the different class of coherences in $\rho_{\rm cl}$ and analyze their effect on the power spectrum of the single qubit probe.
In particular, we present how it is possible 
to shape the spectrum using the so called 
{\em displacement coherences} (DCs)~\cite{Angsar}, which are the matrix elements of 
$\rho_{\rm cl}$ that contribute to the expectation values
$\langle J_\pm\rangle$~\cite{CerenEntropy,BarisThermal}.
Under the assumptions outlined in Sec.~\ref{one_qubit}, 
we can obtain an equation that rules the 
evolution of the single qubit probe 
assuming that only DCs are present in the coherent atomic cluster.
Up to second order in $g\texttt{t}$ the
equation describing the dynamics reads as (see Appendix~\ref{apx_micro} 
for details):
\begin{align}\label{unitary_evolution}
\dot\rho_q=-i[H_{\rm eff},\rho_q],
\end{align}
where
\begin{align}
H_{\rm eff}=\tilde\mu\langle J_+\rangle\sigma_-
+\tilde\mu\langle J_-\rangle \sigma_+
\end{align}
is the effective Hamiltonian.
Note that $H_{\rm eff}$ emulates the 
semi-classical radiation-matter interaction 
between a two-level atomic system and a 
classical electric field. 
It is possible to identify an effective Rabi 
frequency given by
$\Omega_{\rm eff}=2\tilde\mu|\langle J_\pm\rangle|$
with $\tilde\mu=pg\texttt{t}$.
In order to clearly and explicitly present
the action of DCs on the spectrum of the 
qubit probe, we set 
$\gamma=\gamma_\phi=\bar{n}_{\rm en}=0$,
$\langle J_\pm^2\rangle=\langle J_\pm J_\mp\rangle=0$,
and keep only $\langle J_\pm\rangle\neq 0$ which guarantees that atomic bath clusters contain only DCs. 

Since Eq.~(\ref{unitary_evolution}) describes unitary 
evolution of the system, it does not present
a steady-state solution and thus we cannot
make use of the quantum-regression formula
to compute the dipole-field autocorrelation 
function. Therefore, it is not possible to 
calculate the idealized stationary power 
spectrum definition presented in Eq.~(\ref{specdef1}).
Instead, we will use the physical spectrum introduced by 
Eberly and W\'odkiewicz (EW)~\cite{Eberly77}. 
In general, this spectrum definiton can be applied 
to non-stationary light sources. For the
single qubit probe it is defined as:
\begin{align}\label{spec_ebe1}
S(\omega,t,\Gamma_{\rm f})=2\Gamma_{\rm f}&\int_{0}^t dt_1\int_{0}^t dt_2
e^{-(\Gamma_{\rm f}-i\omega)(t-t_1)}\nonumber\\
&\times e^{-(\Gamma_{\rm f}+i\omega)(t-t_2)}
\langle \sigma_{+}(t_1)\sigma_{-}(t_2)\rangle,
\end{align}
where $\omega$ and $\Gamma_{\rm f}$ are the central 
frequency and band half-width of a Fabry-P\'erot 
cavity acting as a filter, respectively, and we assume its central 
line matches the qubit frequency $\omega_q$. 
The EW-spectrum is more realistic, it
reduces to the previously introduced definition by Wiener-Khintchine for the stationary state,
$t\rightarrow\infty$, and in the limit of 
infinite resolution,
$\Gamma_{\rm f}\rightarrow0$~\cite{Eberly77}. 
If the qubit probe is initially in its ground state
the dipole-field auto-correlation function 
is easily obtained [see Eq.~\eqref{time_corre}], and
we can use it to express 
Eq.~\eqref{spec_ebe1} in following way
\begin{align}\label{spec_ebe2}
S(\omega,t,\Gamma_{\rm f})=
&\frac{\Gamma}{2} e^{-2\Gamma_{\rm f} t}\Big{\{}
\Big|\int_{0}^t d\tau e^{(\Gamma_{\rm f}-i\omega)\tau}\sin(\Omega_{\rm eff}\tau)\Big|^2
\nonumber\\+
&
\Big|\int_{0}^t d\tau e^{(\Gamma_{\rm f}-i\omega)\tau}\big[1-\cos(\Omega_{\rm eff}\tau)\big]\Big|^2
\Big{\}}.
\end{align}

Analytical solution for the integrals of 
Eq.~\eqref{spec_ebe2} 
is possible, however, the entire expression of the
full time-dependent spectrum is too cumbersome to show 
here. Instead, a good approximate time-independent 
expression can be obtained in the long-time limit 
($t\gg 1$)~\cite{RicTime}, where after a few Rabi 
oscillations the spectrum has stabilized to three nearby Lorentzians:
\begin{align}\label{spec_ebe_approx}
S(\omega,\Gamma_{\rm f})=&
\frac{\frac{1}{4}\Gamma_{\rm f}}{\Gamma_{\rm f}^2+[(\omega-\omega_q)+\Omega_{\rm eff}]^2}+
\frac{\frac{1}{2}\Gamma_{\rm f}}{\Gamma_{\rm f}^2+(\omega-\omega_q)^2}\nonumber\\
&+\frac{\frac{1}{4}\Gamma_{\rm f}}{\Gamma_{\rm f}^2+[(\omega-\omega_q)-\Omega_{\rm eff}]^2}.
\end{align}
A Mollow-like structure~\cite{MollowPR} can be 
inferred from above expression and its explicit behavior is shown
in Fig.~\ref{spec1}, where the satellite peaks are 
located at $\omega=\omega_q\pm\Omega_{\rm eff}$. Recall that $\Omega_{\rm eff}$ is directly proportional to $\langle J_\pm\rangle$, therefore the magnitude of the DCs defines if the single qubit 
probe is in the weak, moderate or strong 
(where the sidebands emerge) driven regime.
If we add energy losses into our system,
we will be simulating the physical process of 
resonance fluorescence (RF) with an exact Mollow 
spectrum as a final result~\cite{MollowPR}. 
Such a setting can be created within the framework of the model we present in this work. 

\begin{center}
\begin{figure}[t!]
\includegraphics[width=7.8cm, height=5.5cm]{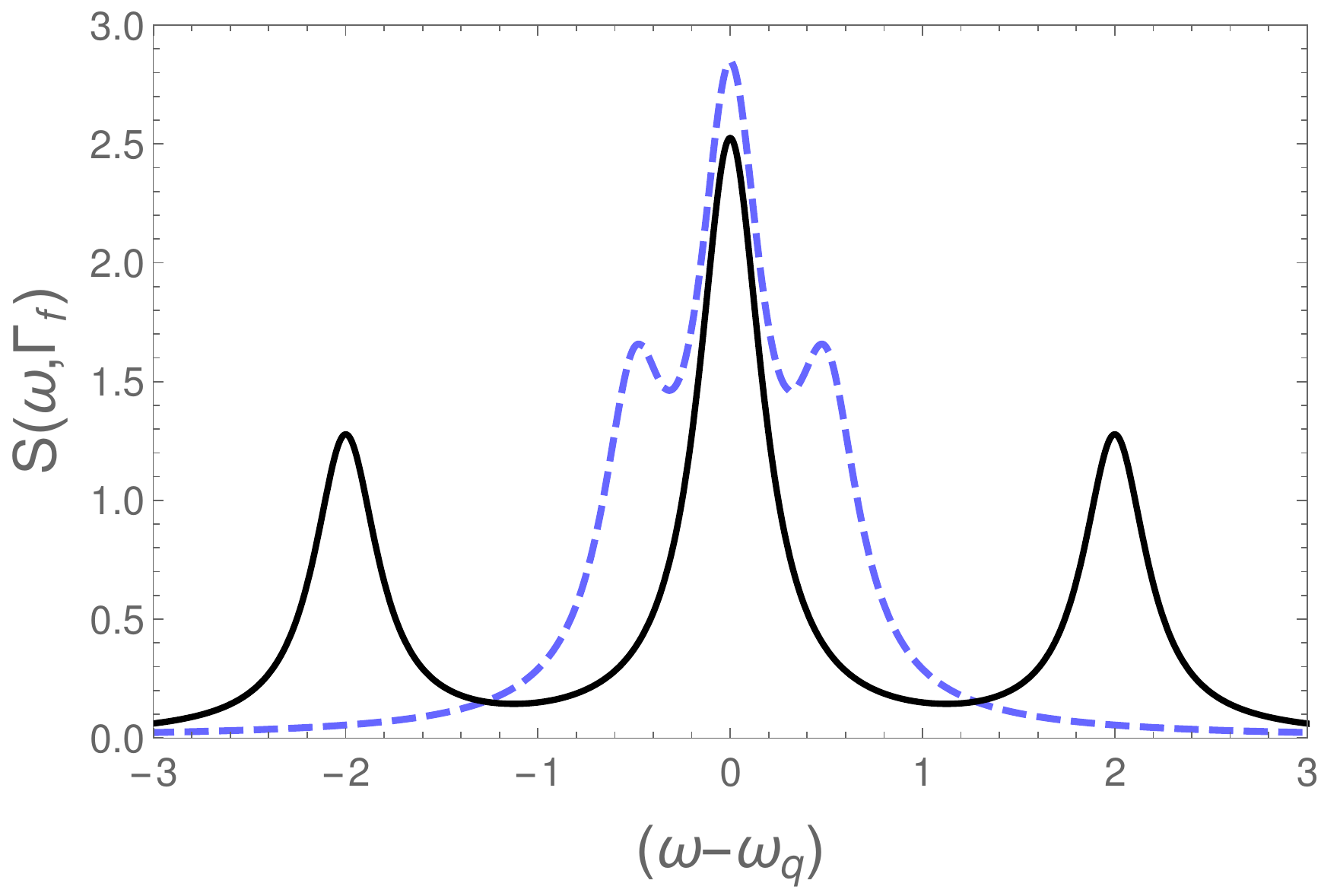} 
\caption{Spectrum the the single qubit
probe~\eqref{spec_ebe_approx} when 
displacement coherences of the $N$-qubit 
cluster are taken into account.
The displacement coherences, $\langle J_\pm\rangle$, 
define the location of the sidebands in this 
Molow-like spectrum. We have set the effective
Rabi frequency 
$\Omega_{\rm eff}\equiv 2\tilde\mu|\langle J_\pm\rangle|$
equals to 2 (black solid line) and 1/2 (blue dashed line).
$\Gamma_{\rm f}=0.2$.\hspace{1.cm}~}
\label{spec1}
\end{figure}
\end{center}

A motivation to study the Mollow spectrum of RF in the suggested configuration, is due to the fact that this has became in a figure of merit for the coherent manipulation of semiconductors quantum dots~\cite{Mollow_QD} and superconducting qubits~\cite{Astafiev840}. We also think that suggested setups are more practical as compared to cavity-QED configurations, in particular, they can be implemented in compact solid-state architectures, where no laser light and/or three-dimensional electromagnetic cavities are need. However, depending on the experimental implementation, unless we prepare the DCs with a well-define phase reference they will cancel out, on average, after each sequential cluster-probe interaction~\cite{Latune_2019}. These operational difficulties can be mitigated by reducing the number of qubits involved in the cluster. Actually, we only need just one qubit with DCs in order to perform the quantum simulation of resonance fluorescence, however, the corresponding Rabi frequency that one can obtain in such configuration will be small one and maybe hard to distinguish from the central peak. Similarly, choosing a cluster with a minimum size of two-qubits, and assuming that all different types of coherences are present in the non-thermal bath, i.e.
$\langle J_\pm\rangle$, $\langle J_\pm^2\rangle$ 
and $\langle J_\pm J_\mp\rangle$ are different 
from zero, \textcolor{black}{then} we could perform the quantum
simulation of RF from a single two-level system
in an artificial squeezed vacuum by a 
repeated interaction scheme. 
This scheme generates an effective Markovian master equation (\ref{apen_micr_mast}) identical to that of a driven two-level atom in a squeezed vacuum in free space, without the necessity to use sources of true squeezed light. As the scheme consists of only qubit-qubit interactions it can provide arbitrarily strong squeezing~\cite{CerenEntropy}. Therefore, this is a viable alternative to control polarization decay and spectral response of a single qubit probe. Notice that we do not have to wait for the steady state solution driven by a large number of repeated interactions or collisions. Equation~(\ref{spec_ebe1}) is valid, in general, for non-stationary sources of light. This means that with just a few number of collisions between the quantum probe and the small coherent clusters, which means a short evolution, the time-dependent spectrum could be, in principle, measured. Actually, in~\cite{RicTime} one of us proved that after just a few Rabi oscillations, a well-define Mollow-like spectrum can be obtained using Eq.~(\ref{spec_ebe1}).

In addition to spectrum engineering and longevity of quantum coherence, these results can be significant for spectral characterization of unknown quantum resources for their quantum information and energy resource values as well. From the quantum thermodynamic point of view, 
the emergence of a Mollow-like triplet in the 
spectroscopy of the corresponding qubit probe 
should be considered as a strong signature of the system acting as a thermo-mechanical engine, such that the interaction with the non-thermal bath result in both heat and work to be imparted to the probe~\cite{CerenEntropy}.

\section{Experimental relevance}\label{sec:experiment} 

In this section, we would like to elaborate on the experimental side of our motivation to use the power spectrum to determine the apparent temperature of the quantum probe and, ultimately, the apparent temperature of the non-thermal bath.

As we have discussed in Sec.s~\ref{one_qubit} and \ref{one_cavity}, the apparent temperature of the probe can be determined by the excited state population and the mean number of photons at the steady state, respectively. Extracting this information from the power spectra of the probes are in fact possible by looking at the integral of the spectra over all frequencies, i.e. their intensities, as shown in Sec.~\ref{sec:spectrum}. With this knowledge at hand, there are two relevant experimental works performing the so called {\em fluorescence thermometry} of thermal baths using a single quantum dot~\cite{Haupt} and nitrogen-vacancy (NV) centres in diamond nanocrystals (nanodiamonds)~\cite{Kuckso}. In both cases the fluorescence of the quantum thermometer is converted to population measurements, which is in line with the discussion we made above. Apart from the fact that we study quantum thermometry of non-thermal baths, one of the main differences between Ref.s~\cite{Haupt,Kuckso} and our work is that we are considering only spinless quantum probes. For example, in~\cite{Haupt} the ultra-cold temperature of the quantum dot is extracted by combining, the Fermi-Dirac distribution and the relative amplitudes of the absorption resonance for the high-energy and low-energy optical transitions (see Eq.~(1) of~\cite{Haupt}). Moreover, in quite contrast with our work, the precise value of the transition frequency between the ground state and the two-degenerate excited states of the NV-centres, has a temperature dependence due to thermally induced lattice strains. Therefore, the operational principle of this quantum thermometer relies on the accurate measurement of such transition frequency, which can be done with high spatial nanometer resolution~\cite{Kuckso}. In our work, the corresponding transition frequencies of the quantum probes do not depend on the temperature.

Regarding the issue of which technique could be used experimentally to determine the probe power spectrum, we would like to emphasize that our results are general and depending on the specific experimental implementation of the setup, the power spectrum should be measured in the appropriate context. For instance, if our scheme is implemented in a circuit-QED architecture, one should record the resonator transmission spectrum using a continuous-wave measurement at low drive~\cite{Blais}, where the transmission frequencies are spectrally resolved in a heterodyne detection scheme~\cite{Fink2008}. Actually, in the linear response limit the drive term can be neglected from the coherent dynamics~\cite{Fink2010}, which is really useful for the numerical calculation of the transmission spectrum. On the other hand, if our scheme is implemented in a cavity-QED setup, the fluorescence spectrum of the quantum probe could be measured using, for example, the standard balance homodyne detection schemes~\cite{castro2016}. In this technique the phase of the unknown signal is measured by mixing it with a strong local oscillator (LO) in a beam splitter, where the LO is a field with a well-define amplitude and phase. Using a spectrum analyzer, the output intensity from each beam splitter ports are subtracted electronically and the desired signal is obtained as a result, which is proportional to the amplitude of the quantum field~\cite{gerry_knight_2004}.

Another field of research that experimental assessment of the apparent temperature of quantum systems is highly relevant are quantum Hall systems. The topological order of some fractional quantum Hall states are better characterized by the thermal Hall conductance, which is a topological invariant and therefore has quantized values \cite{Banerjee2017,Banerjee2018,Ma2019}. It is possible to distinguish between Abelian and non-Abelian states of matter by identifying if the thermal Hall conductance is quantized to integer and non-integer values, respectively. The latter case opens up the possibility of using these systems for braiding of non-Abelian particles which is beneficial for topological quantum computing. However, precise determination of thermal Hall conductance requires precise information on the temperature of the state. We hope that the scheme presented in the present work can contribute along these lines.

\section{Conclusions}\label{sec:conclusions}

We have studied the thermalization dynamics of two different types of quantum probes that are in contact with a thermal and non-thermal, coherent bath. We have modeled our quantum probes as a single-qubit and a single-mode cavity field, and showed the non-trivial effects of the coherences contained in the non-thermal bath (HECs) on the thermalization temperature and time of these model systems. We have proposed a strategy to measure these effects by investigating the power spectrum of the probes in both cases, and proved that the spectral bandwidth and intensity are explicitly dependent on HECs. Therefore, the suggested method is capable of identifying the apparent temperature of a non-thermal bath with HECs, as well as the temperature of a thermal bath, in spectroscopic experiments. We think that these results also contribute to the field of quantum thermometry which aims to estimate the temperature of an environment using a quantum probe. The presented spectroscopic method is not only capable of assessing the temperature of a thermal bath, but also points a direction on how to identify the apparent temperature (positive and negative) that a quantum probe sees when in contact with an non-thermal environment with only thermalizing coherences. Through such an apparent temperature, quantum thermometry of multipartite coherences and correlations could be possible for non-thermal environments. Another application of this method is the implementation of quantum simulation of resonance fluorescence using different types of coherences in the non-thermal bath. On top of being an significant result per se, observation of such an effect can again serve as a tool for characterizing an unknown environment via a single-qubit probe. Even though the presented repeated interaction scheme may present some practical challenges at the implementation stage, for example the preparation or re-initialization of the coherent cluster, we hope that the results presented in this work can further induce investigations on non-thermal baths through spectral measurements of a probe system.

\begin{acknowledgments}
\"O. E. M. acknowledges support by TUBITAK 
(Grant No. 116F303) and by the EU-COST Action (CA15220).
\end{acknowledgments}

\appendix

\section{Generalized micromaser like equation}\label{apx_micro}
We make the derivation, following closely
Ref.~\cite{singlemachine}, of the micromaser 
master like equation used in Eqs.~\eqref{master1}
and~\eqref{mastercavity} of the main text.
We start by considering a ``multipulse"
type interaction between a central system $S$ 
and a bath system $B$ that is described by the 
linear coupling Hamiltonian
$H_{I}=\epsilon(B^\dagger s+Bs^\dagger)$,
where $s, s^\dagger$ ($B, B^\dagger$) are the
annihilation and creation, central (bath) system 
operators respectively;
$\epsilon$ is the coupling coefficient. 
The corresponding evolution operator is 
$U(\tau)=\exp(-i H_{I}\tau)$.
Before each interaction at time $t_j$, it is 
assumed that the state of the bath system $B$ 
is reset to its initial value which does 
not have any correlation with $S$, so the 
total state is 
$\rho(t_j)=\rho_\text{\tiny S}(t_j)\otimes\rho_\text{\tiny B}(0)$.
This assumption seems arbitrary, but actually, 
it is a typical physical situation in one-atom 
masers~\cite{Filipowicz} for which, individual 
and independent Rydberg atoms pass one by one 
through a high-finesse electromagnetic 
cavity~\cite{Meschede,RempeMaser,RempeMaser2}.
After each interaction of duration 
$\tau$, the central system density matrix is
$\rho_\text{\tiny S}(t_j+\tau)=\text{Tr}_\text{\tiny B}
\{U(\tau)\rho(t_j)U^\dagger(\tau)\}$.
If we introduce a rate $p$ of a Poisson process 
to portray this interaction, then
in a time interval $(t+\delta t)$ the 
probability of interaction will be $p\delta t$.
As a consequence the state of the central system 
$S$ can be written during this time interval as 
the sum of the two possible outcomes
$\rho_\text{\tiny S}(t+\delta t)=p\delta t\text{Tr}_\text{\tiny B}
{\{}U(\tau)\rho(t)U^\dagger(\tau){\}}+
(1-p\delta t)\rho_\text{\tiny S}(t)$,
where $1-p\delta t$ is the probability of 
having a non interaction event. 
If we divided the above equation by $\delta t$ 
and take the limit when it goes to zero, we obtain 
the following differential equation for the central 
system density matrix
\begin{align}\label{apx_mast}
\dot\rho_\text{\tiny S}(t)=p\big[{\rm Tr}_\text{\tiny B}\{U(\tau)
\rho_\text{\tiny S}(t)\otimes\rho_\text{\tiny B}(0) U^\dagger
(\tau)\}-\rho_\text{\tiny S}(t)\big],
\end{align}
where we have used the derivative definition:
$\text{d}\rho_\text{\tiny S}(t)/\text{d}t\equiv\lim_{\delta t\rightarrow 0}
\left(\rho_\text{\tiny S}(t+\delta t)-\rho_\text{\tiny S}(t)\right)/\delta t$.
By now~\eqref{apx_mast} is valid for any coupling
strength and interaction time, however, in 
micromaser setups the product $\epsilon\tau$ is
normally small~\cite{singlemachine}. 
We proceed to approximate the evolution operator 
up second order in $\epsilon\tau$ as:
$U(\tau)\sim 1-U_{1}(\tau)-U_{2}(\tau)$,
where
\begin{align}
U_{1}(\tau)&=i\epsilon\tau(B^\dagger s+B s^\dagger),\label{u1}\\
U_{2}(\tau)&={(\epsilon\tau)}^2(B^\dagger s+B s^\dagger)^2/2.\label{u2}
\end{align}
Inserting~\eqref{u1}, \eqref{u2}
in~\eqref{apx_mast}, it yields
\begin{align}\label{ap_a2}
\dot\rho_\text{\tiny S}(t)=p&{\rm Tr}_\text{\tiny B}\big{\{}
U_1(\tau)\rho(t)U_1^\dagger(\tau)-U_1(\tau)\rho(t)\nonumber\\
&-\rho(t)U_1^\dagger(\tau)
-U_2(\tau)\rho(t)-\rho(t)U_2^\dagger(\tau)\big{\}},
\end{align}
where we have kept only terms up second order 
in $\epsilon\tau$ neglecting
$U_1(\tau)\rho(t)U_2^\dagger(\tau)\propto(\epsilon\tau)^3$,
$U_2(\tau)\rho(t)U_2^\dagger(\tau)\propto(\epsilon\tau)^4$
and their complex conjugate\revision{s}. 
Hereafter we denote $\rho_\text{\tiny S}(t)$ as $\rho_\text{\tiny S}$.
Tracing out of the bath degrees of freedom 
in~\eqref{ap_a2} we obtain the following 
micromaser master like equation
\begin{align}\label{apen_micr_mast}
\dot\rho_\text{\tiny S}=-i&[H_{\rm eff},\rho_\text{\tiny S}]+
\frac{\mu}{2}\langle B B^\dagger\rangle\mathcal{L}[s]\rho_\text{\tiny S}+
\frac{\mu}{2}\langle B^\dagger B\rangle\mathcal{L}[s^\dagger]\rho_\text{\tiny S} \nonumber\\+
&\frac{\mu}{2}\langle B^2\rangle\mathcal{L}_\text{{\tiny S}q}[s^\dagger]\rho_\text{\tiny S}+
\frac{\mu}{2}\langle B^{\dagger 2}\rangle\mathcal{L}_\text{{\tiny S}q}[s]\rho_\text{\tiny S},
\end{align}
where $\mu=p(\epsilon\tau)^2$ and 
$H_{\rm eff}=p\epsilon\tau(\langle B^\dagger\rangle s+\langle B\rangle s^\dagger)$.
The average values are taken with respect to
the initial bath state, 
i.e., 
$\langle \mathcal{O}\rangle={\rm Tr}\{\rho_\text{\tiny B}(0)\mathcal{O}\}$.
The Lindblad superoperators are defined as:
$\mathcal{L}[o]\rho_\text{\tiny S}\equiv 2o\rho_\text{\tiny S}{o^\dagger}
-{o^\dagger}{o}\rho_\text{\tiny S}-\rho_\text{\tiny S}{o^\dagger}{o}$
and
$\mathcal{L}_\text{{\tiny S}q}[{o}]\rho_\text{\tiny S}\equiv 2{o}\rho_\text{\tiny S}{o}
-{o}^2\rho_\text{\tiny S}-\rho_\text{\tiny S}{o}^2$.
During the derivation of~\eqref{apen_micr_mast}
it was not necessary to specify if system $S$ 
and/or $B$ were from a bosonic or fermionic nature. 
Moreover, no restrictions have been made regarding 
their composition, $S$ and $B$ can represent either 
individual quantum systems or 
several non-interacting ones.
As long as their interaction Hamiltonian can be 
written in the form of $H_I$ together with the previous assumptions,
\eqref{apen_micr_mast} will be valid.

It was assumed that between each interaction the central system $S$ evolves unitarily.
However, it is possible that $S$ could experience 
energy losses and/or decoherence during the evolution 
due to the unavoidable interaction with its surrounding. 
In such a case, it easy to prove 
(see Ref.~\cite{singlemachine}),
within the approach of the theory of open quantum
systems~\cite{carmichael}, 
that the only modification to~\eqref{apen_micr_mast}
is  to add another series of Lindbladians, 
$\propto\mathcal{L}_j[\mathcal{O}_j]\rho_\text{\tiny S}$, 
for each $j$ dissipation and/or dephasing process.

Equation~\eqref{apen_micr_mast} reduces to 
Eq.~\eqref{master1} [Eq.~\eqref{mastercavity}]
choosing $H_I$ as $H_\text{dip}$ ($H_\text{TC}$).
In such case bath system $B$ would be the 
non interacting $N$-qubit coherent cluster and
we replace $B$ by $J_-$.  Central system $S$ 
will be the single-qubit (single cavity field mode) 
so that $s\rightarrow \sigma$ ($a$) and 
$\mu\rightarrow \mu_q$ ($\mu_c$). 
Additionally, we demand that
$\rho_\text{\tiny B}(0)$ satisfy 
$\langle B\rangle=\langle B^2\rangle=0$
while $\langle B^\dagger B\rangle$ and
$\langle BB^\dagger\rangle$ should be different 
from zero. On the other hand, if we allow only
$\langle B\rangle$ and $\langle B^\dagger\rangle$ to
survive, just the unitary part of
\eqref{apen_micr_mast} will remain,
this is the condition to derive 
Eq.~(\ref{unitary_evolution}).

Eq.~\eqref{apen_micr_mast} is quite general and
it is applicable to several micromaser based 
models \cite{CerenEntropy,BarisThermal,gerzontemp}
including superra\-diant systems~\cite{hardal} and
entangled qubit environments~\cite{Shakib}. 
Also it works for the case of gene\-ra\-li\-zed 
quantum ``phaseonium fuels"~\cite{DenizLevels}, 
\revision{where} the bath system consists of a $N+1$ level
atom with $N$ degenerate coherent ground 
states~\cite{DenizTime}.
Unlike those previous works, \eqref{apen_micr_mast} 
allows to get, more easily, analytical results.

\section{Autocorrelation functions}\label{apx_corr}

In the stationary case we know that
$\langle \sigma_+(0)\sigma_-(\tau)\rangle$=
$\lim_{t\rightarrow\infty}\langle \sigma_+(t)\sigma_-(t+\tau)\rangle$
is the first order dipole-field 
(${E}^{(+)}\propto\sigma_-$) autocorrelation 
function in  the steady state. 
The quantum regression formula (QRF) allows us
to obtain such two-time correlation from the 
single-time function
$\langle \sigma_-(t)\rangle$~\cite{carmichael}. 
Identifying that $\langle \sigma_-(t)\rangle=\rho_{eg}(t)$
and $\langle \sigma_+(t)\sigma_-(t)\rangle=\rho_{ee}(t)$, 
the use of the QRF in Eq.~\eqref{correl} gives
$\langle \sigma_+(t)\sigma_-(t$+$\tau)\rangle
=\rho_{ee}(t)\exp[-(2\gamma_\phi+(2t_q)^{-1})\tau]$;
which in the steady state limit and in the 
Schr\"odinger picture yields
\begin{equation}
\langle \sigma_+(0)\sigma_-(\tau)\rangle
=\rho_{ee}^{st}\exp[-(i\omega_q+2\gamma_\phi+(2t_q)^{-1})\tau].
\end{equation}

For the time-dependent case, the solution for
the density matrix is 
$\rho_q(t)=U_q(t)\rho_q(0)U_q^\dagger(t)$,
where
$U_q(t)=\cos(t\Omega_{\rm eff}/2)-i\sigma_x\sin(t\Omega_{\rm eff}/2)$
~\cite{MayaApplied}.
Using $U_q(t)$ the dipole-field autocorrelation 
function for the qubit, starting in its ground state
$\rho_q(0)=|g\rangle \langle g|$, is
\begin{align}\label{time_corre}
\langle g|\sigma_+(t_1)\sigma_-(t_2)|g\rangle=&
\sin^2\big(t_1	\Omega_{\rm eff}/2\big)\sin^2\big(t_2\Omega_{\rm eff}/2\big)
\nonumber\\+&
\sin\big(t_1\Omega_{\rm eff}\big)\sin\big(t_2\Omega_{\rm eff}\big)/4,
\end{align}
which we use in the integral of Eq.~\eqref{spec_ebe1}
to obtain Eq.~\eqref{spec_ebe2}.

\bibliography{references}

\end{document}